\documentclass[a4style,aip,cha,tightenlines,amsmath,amssymb,amsfonts,showpacs]{revtex4}

\usepackage[english]{babel}
\usepackage{times}
\usepackage{color}
\usepackage{oldgerm,amsmath,pifont,amsfonts,amssymb,latexsym}
\usepackage{float,xspace,calc,ifthen}
\usepackage{enumerate,psboxit}
\usepackage[dvips]{epsfig,graphicx}
\usepackage{hyperref}
\hypersetup{
    colorlinks,%
    citecolor=blue,%
    filecolor=blue,%
    linkcolor=blue,%
    urlcolor=blue
}
\usepackage{changebar}
\usepackage{verbatim}
\usepackage{amsthm}

\usepackage{textcomp}
  \expandafter\let\csname equation*\endcsname\relax
  \expandafter\let\csname endequation*\endcsname\relax

\theoremstyle{plain}
\newtheorem{thm}{Theorem}

\newtheorem{prop}[thm]{Proposition}

\newtheorem{oss}{Remark}

\newcommand{\Z}{\mathbb{Z}}
\newcommand{\N}{\mathbb{N}}
\newcommand{\hX}{\widehat{X}}

\begin{document}

\title[Non-chaotic map and anomalous diffusion]{A simple
non-chaotic map generating subdiffusive, diffusive and superdiffusive
dynamics}

\author{Lucia Salari} 
\email{lucia.salari@polito.it}
\affiliation{Dipartimento di Scienze Matematiche, Politecnico di Torino,
Corso Duca degli Abruzzi 24 I-10129 Torino, Italy}

\author{Lamberto Rondoni} 
\email{lamberto.rondoni@polito.it}
\affiliation{Dipartimento di Scienze Matematiche, Politecnico di Torino,
Corso Duca degli Abruzzi 24 I-10129 Torino, Italy}
\affiliation{Graphene@PoliTO Lab, Politecnico di Torino, Corso Duca degli Abruzzi 24 I-10129 Torino, Italy}
\affiliation{INFN Sezione di Torino, Via P. Giuria 1, 10125 Torino, Italy}

\author{Claudio Giberti}
\email{claudio.giberti@unimore.it}
\affiliation{Dipartimento di Scienze e Metodi dell'Ingegneria, Universita' di Modena e Reggio E.,
Via G. Amendola 2 - Pad. Morselli, I-42122 Reggio E., Italy}

\author{Rainer Klages}
\email{r.klages@qmul.ac.uk}
\affiliation{School of Mathematical Sciences, Queen Mary University of London, Mile End Road, London E1 4NS, UK}

\begin{abstract} 
  Analytically tractable dynamical systems exhibiting a whole range of
  normal and anomalous deterministic diffusion are rare. Here we
  introduce a simple non-chaotic model in terms of an interval
    exchange transformation suitably lifted onto the whole real line
  which preserves distances except at a countable set of points. This
  property, which leads to vanishing Lyapunov exponents, is designed
  to mimic diffusion in non-chaotic polygonal billiards that give rise
  to normal and anomalous diffusion in a fully deterministic
  setting. As these billiards are typically too complicated to be
  analyzed from first principles, simplified models are needed to
  identify the minimal ingredients generating the different transport
  regimes. For our model, which we call the slicer map, we calculate
  all its moments in position analytically under variation of a single
  control parameter.  We show that the slicer map exhibits a
  transition from subdiffusion over normal diffusion to superdiffusion
  under parameter variation.  Our results may help to understand the
  delicate parameter dependence of the type of diffusion generated by
  polygonal billiards.  We argue that in different parameter regions
  the transport properties of our simple model match to different
  classes of known stochastic processes. This may shed light on
  difficulties to match diffusion in polygonal billiards to a single
  anomalous stochastic process.\\ \\

\end{abstract} 


\date{\today}

\maketitle


\begin{quotation} 

  \bf Consider equations of motion that generate dispersion of an
  ensemble of particles as the dynamics evolves in time. A fundamental
  challenge is to develop a theory for predicting the diffusive
  properties of such a system starting from first principles, that is,
  by analyzing the microscopic deterministic dynamics. Here we
  introduce a seemingly trivial toy model that, analogously to
  polygonal billiards, exhibits dispersion but is not chaotic in
  terms of exponential sensitivity with respect to initial
  conditions. We show that our simple map model generates a
  surprisingly non-trivial spectrum of different diffusive properties
  under parameter variation.

\end{quotation}

\section{Introduction}

How macroscopic transport emerges from microscopic equations of motion
is a key  topic in dynamical system theory and nonequilibrium statistical
  physics \cite{Do99,Gasp,Kla06,RoMM07,CFLV08,BPRV08,JR10}. While microscopic
  chaos, characterized by positive Lyapunov exponents, typically leads
  to Brownian motion-like dynamics by reproducing conventional
  statistical physical transport laws, for weakly chaotic dynamical
  systems where the largest Lyapunov exponent is zero the situation
  becomes much more complicated \cite{Zas02,Kla06,KRS08}. Such
  non-trivial dynamics is relevant for many topical
  applications like, for example, nanoporous transport
  \cite{JBS03,IRP,JeRo06,JBR08,Sokolov}. In the
  former case the mean square displacement (MSD) of an ensemble of
  particles grows linearly in the long time limit, $\langle
  x^2\rangle\sim t^{\gamma}$ with $\gamma=1$ defining normal
  diffusion. In the latter case one typically
  finds anomalous diffusion with $\gamma\neq1$, where for $\gamma<1$
  one speaks of subdiffusion, for $\gamma>1$ of superdiffusion
  \cite{MeKl00,Zas02,KRS08,Sokolov}.

\begin{figure}
\centering
\includegraphics[scale=0.7,angle=-90]{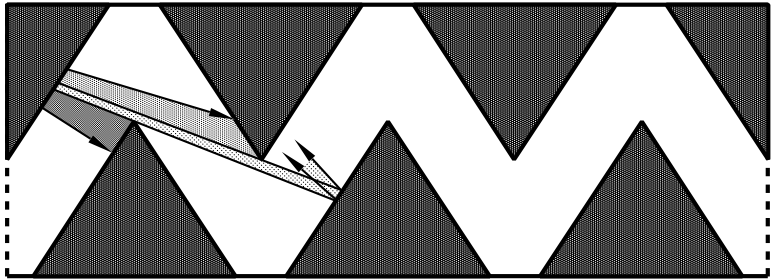}
\caption{Example of a polygonal billiard channel in which single point
  particles scatter elastically with sawtooth walls
\cite{LWH02,LWWZ05,Kla06,JeRo06}. Shown is how a beam of
  particles is split by the corners of the billiard while
  propagating.}
  \label{fig:poly_chann} 
\end{figure}

To our knowledge only a few deterministic dynamical systems are
  known exhibiting all three regimes of subdiffusion, normal diffusion
  and superdiffusion under parameter variation. Examples of
  one-dimensional maps are a Pomeau-Manneville like model where
  anomalous diffusion originates from an interplay between
  different marginally unstable fixed points \cite{ZuKl95}. The
  climbing sine  map displays exactly three different diffusive regimes with
  $\gamma=0,1,2$ corresponding to periodic windows and chaotic regions
  connected to period doubling bifurcations and crises
  \cite{KoKl02,KoKl03}. For the two-dimensional standard map numerical
  evidence exists for a transition from sub- to superdiffusion
  generated by a mixed phase space \cite{ThaRo14}. Least understood is
  diffusion in two-dimensional polygonal billiards
  \cite{Zas02,LWH02,LWWZ05,Kla06,Dett14,JeRo06,JBR08}, see Fig.~\ref{fig:poly_chann} for
  an example. By definition these systems exhibit linear dispersion of
  nearby trajectories with zero Lyapunov exponents for typical initial
  conditions, hence non-chaotic behavior. However, they nevertheless
  generate highly non-trivial dynamics due to complicated topologies
  yielding pseudohyperbolic fixed points and pseudointegrability. For
  this reason they are sometimes called {\em
    pseudochaotic} \cite{Zas02,Kla06}. A line of numerical work on
  periodic polygonal billiard channels revealed sub-, super- and
  normal diffusion depending on parameter variation
  \cite{ArGuRe00,ARV03,SchSt06,JeRo06,SaLa06,JBR08}. Rigorous analytical
  results are so far only available for periodic wind-tree models
  supporting an extremely delicate dependence of diffusive properties
  on variation of control parameters \cite{HaWe80,DHL11pp}. For the
  mathematical derivations it has been exploited that polygonal billiards
  can often be reduced to interval exchange transformations (IETs),
  see also \cite{HaMc90,Gut96,ZaEd01}. These are one-dimensional maps
  generalising circle rotations which cut the original interval into
  several subintervals by permuting them non-chaotically. Both
  polygonal billiards and IETs are known to exhibit hihgly non-trivial
  ergodic properties, and in general there does not seem to exist any
  theory to understand the complicated diffusive dynamics of such
  systems from first principles. Random non-overlapping wind-tree
  models and related maps, on the other hand, enjoy a kind of
stochasticity which appears analogous to the dynamically generated randomness
of chaotic systems, leading to sufficiently rapid decay of correlations
and good statistical properties. Consequently, these models have been found to
  yield normal diffusion that is indistinguishable from Brownian
  motion \cite{DC00,CFVN02}. In contrast, periodic polygonal billiards
  have very long lasting dynamical correlations and poor statistical
  properties, which are associated with very sensitive dependence of
  their transport properties on the details of their geometry
  \cite{JeRo06}.

These difficulties to
  understand diffusion in polygonal billiards on the basis of
  dynamical systems theory are paralleled by difficulties in attempts
  to approximate their diffusive properties by stochastic theory:
  There still appears to be a controversy in the literature of whether
  continuous time random walk theory and L\'evy walks, fractional
  Fokker-Planck equations or scaling arguments should be applied to
  undestand their anomalous diffusive properties, with different
  approaches yielding different results for the above exponent
  $\gamma$ of the MSD \cite{Zas02,DKU03,LWWZ05,Kla06}. While all
  these theories are based on dynamics generated from temporal randomness,
  spatial randomness leads yet to another important class of stochastic
  models, called random walks in random environments, which yields
  related types of anomalous diffusion: An important example in one
  dimension is the L\'evy Lorentz gas where the scatterers are
  randomly distributed according to a L\'evy-stable probability
  distribution of the scatterer positions. This model has been
  studied both numerically and analytically revealing a highly
  non-trivial superdiffusive dynamics that depends in an intricate way
  on initial conditions and the type of averaging
  \cite{BF99,BFK00,BCV10}. This work is related to experiments on
  L\'evy glasses where similar behavior has been observed
  \cite{BBW08}.

\begin{figure}
\centering
\includegraphics[scale=0.6]{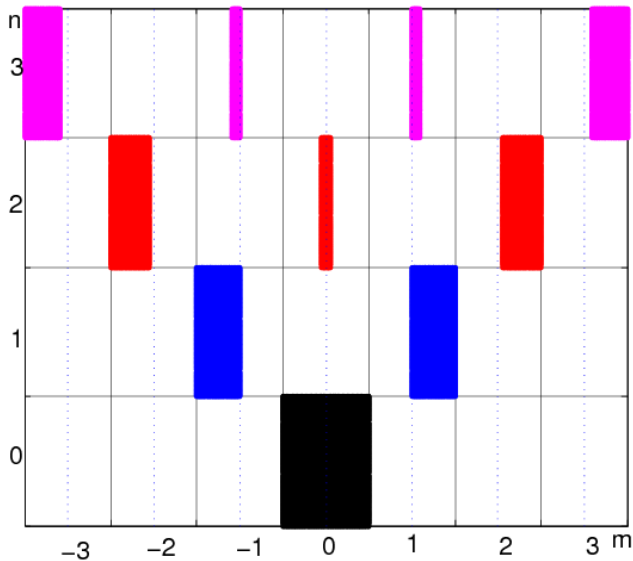}
\caption{Space-time plot illustrating the action of the
  one-dimensional slicer map $S_{\alpha}$ for
  $\alpha=1/3$ 
  defined by Eqs.~(\ref{eq:MAPPAGEN},\ref{eq:SLICEGEN}), where $m$
  (horizontal axis) denotes space and $n$ (vertical axis) time: Shown
  is the diffusive spreading of 
  points that at $n=0$ are uniformly distributed on the unit interval
  centered around $m=0$. As the map is one-dimensional the columns are
  only a guide to the eye. This map is designed to mimic the mechanism
  of beam-splitting in polygonal billiards depicted in
  Fig.~\ref{fig:poly_chann}.}
  \label{fig:slicer} 
\end{figure}

Motivated by the problem of understanding diffusion in polygonal
  billiards, in this paper we propose a seemingly trivial non-chaotic
  map by which we attempt to mimic the dynamics illustrated in
  Fig.~\ref{fig:poly_chann}: Shown is a beam of point particles and
  how it splits due to the collisions at the singularities (corners)
  of a polygonal billiard channel. This mechanism is intimately
  related to the connection between polygonal billiards and IETs
  referred to above. We thus try to capture this slicing dynamics by
  introducing a specific IET defined on a one-dimensional lattice, see
  Fig.~\ref{fig:slicer}. Here the loss of particles
  propagating further in one direction is modeled by introducing
  a deterministic rule following a power law for the jumps from unit
  cell to unit cell.

This simple non-chaotic
  model, which we call the slicer map, generates a surprisingly
  rich spectrum of diffusive dynamics under parameter variation that
  includes all the different diffusion types mentioned above. We mention in passing
  that it provides another example where normal diffusion is
  obtained from non-chaotic dynamics. However, differently from the
  cases of \cite{DC00,CFVN02} and analogously to periodic polygonal
  billiards, it is completely free of randomness. Our simple
  model might help to understand why the type of diffusion in
  polygonal billiards is so sensitive under parameter variation. It
  might also shed some light on the origin of the difficulty to model
  polygonal billiard dynamics as a simple stochastic process.

Our paper is organized as follows: In Section 2 we define the slicer
model and analytically calculate its diffusive properties. To the end
of this section we study our model analytically and illustrate it
numerically for a parameter value that is characteristic for the
dynamics.  In Section 3 we compare the deterministic slicer dynamics
with existing stochastic models of anomalous diffusion. Section 4
contains concluding remarks.

\section{The slicer dynamics}
\subsection{Theory}
Consider the unit interval $M:=[0,1]$, the chain of such intervals
$\widehat{M}:=M\times \mathbb{Z}$, and the product measure
$\hat{\mu}:=\lambda\times\delta_{\mathbb{Z}}$ on $\widehat{M}$, where
$\lambda$ is the Lebesgue measure on $M$ and $\delta_{\mathbb{Z}}$ is
the Dirac measure on the integers.  Denote by $\pi_{M}$ and
$\pi_{\mathbb{Z}}$ the projections of $\widehat{M}$ on its first and
second factors.  Let $x$ be a point in $M$, $\hX=(x,m)$ a point in
$\widehat{M}$, and $\widehat{M}_{m}:=M\times\left\{m\right\}$ the
$m$-th cell of $\widehat{M}$. Subdivide each $\widehat{M}_{m}$ in four
sub-intervals, separated by three points called {\em ``slicers''},
$$\left\{ 1/2 \right\}\times\left\{m\right\}~, \quad
\{\ell_m\}\times\left\{m\right\}~, \quad 
\{1-\ell_m\}\times\left\{m\right\}~, 
$$ 
where $0 < \ell_m < 1/2$ for every $m \in \mathbb{Z}$. 

The slicer model is the dynamical system $(\widehat{M},\hat{\mu},S)$
which, at each time step $n\in\mathbb{N}$, moves all sub-intervals
from their cells to neighbouring cells, implementing the rule $S :
\widehat{M} \to \widehat{M}$ defined by
\begin{equation}\label{eq:MAPPAGEN}
S(x,m)= 
\left\{
\begin{array}{rl}
(x,m-1) & \mbox{ if } 0\leq x <  \ell_{m} \mbox{ or } \frac{1}{2} < x \leq 1-\ell_{m},\\
(x,m+1) & \mbox{ if } \ell_{m} \le x \leq \frac{1}{2} \mbox{ or } 1-\ell_{m} < x \leq 1.
\end{array}
\right.
\end{equation}
For every $\alpha > 0$, let us introduce the family of slicers
\begin{equation} \label{eq:SLICEGEN}
L_{\alpha}=\left \{\Big(\ell_{j}(\alpha), 1-\ell_j(\alpha)\Big) : 
\ell_j(\alpha) = \frac{1}{\left(\left|j\right|+2^{1/\alpha}\right)^{\alpha}},\, j\in \mathbb{Z}\right\}\quad .
\end{equation}
The slicer map is denoted by $S_\alpha$ if all slicers of
Eq.(\ref{eq:MAPPAGEN}) belong to $L_\alpha$:
$\ell_m=\ell_m(\alpha)$. Obviously, for every $\alpha>0$, $S_\alpha$
preserves $\hat{\mu}$ and is not chaotic: Its Lyaponuv exponent
vanishes, as different points in $\widehat M$ neither converge nor
diverge from each other in time, except when separated by a slicer in
which case their distance jumps. This is like for two particles in a
polygonal billiard where one of them hits a corner of the polygon
while the other continues its free flight, see
Fig.~\ref{fig:poly_chann}. But the separation points constitute a set
of zero $\hat{\mu}$ measure, hence they do not produce positive
Lyapunov exponents.

The dependence of the dynamical rule Eq.(\ref{eq:MAPPAGEN}) on
  the coarse grained position in space $m$ is a crucial aspect of the
  slicer model, which distinguishes it from ordinary IETs. That the
  slicers get closer and closer to the boundaries of the cells when
  the absolute value of $m$ grows is meant to reproduce, in
  a one-dimensional setting, what is illustrated in
  Fig.~\ref{fig:poly_chann}: The corners of periodic polygonal
  billiards split beams of particles into thinner and thinner beams as
  they travel further and further away from their initial cell. In
  two dimensions this operation is fostered by the
  rotations of the beams of particles, something that is not possible
  in a single dimension. This thinning mechanism due to slicing is
  mimicked by the power law dependence in Eq.(\ref{eq:SLICEGEN}). In
  effect this means that our slicer particles perform a deterministic
  walk in a L\'evy potential. This quite trivial setting has, as we
  shall see, rather non-trivial consequences. The power law dependence
  is a mere assumption at this point in order to define our model. It
  would have to be developed further to move the slicer map closer to actual
  polygonal billiard dynamics.

The diffusive properties of the slicer dynamics will be examined by
taking an ensemble of points $\widehat E_0$ in the central cell
$\widehat{M}_0=M\times \left\{0\right\}$ and studying the way
$S_{\alpha}$ spreads them in $\widehat{M}$. One finds that in $n$ time
steps the points of $\widehat E_0$ reach $\widehat{M}_n$ and
$\widehat{M}_{-n}$, and that the cells occupied at time $n$ have odd
index if $n$ is odd, and have even index if $n$ is even. 

More precisely, taking 
\begin{equation}\label{eq:PnDn}
P_{n}=\{ j\in \Z :  j\  \mbox{is even and\ }  |j|\le n\},\quad 
D_{n}=\{ j\in \Z :  j\  \mbox{is odd and\ }  |j|\le n\}
\end{equation}
we have
\begin{equation}
S_{\alpha}^{n}~\widehat{M}_{0}=\bigcup_{j\in P_{n}} \big(R_{j}\times \{j\}\big) \quad \mbox{if~} n \mbox{~is even}~,
\qquad 
S_{\alpha}^{n}~\widehat{M}_{0}=\bigcup_{j\in D_{n}} \big(R_{j}\times \{j\}\big),\quad \mbox{if~} n \mbox{~ is odd}
\end{equation}
where $R_{j}\times \{j\}\subset \widehat{M}_{j}$, and $R_{j} \subset M$ is an interval or a union of intervals if 
$\widehat{E}_{0}=\widehat{M}_{0}$, with $R_i \cap R_j = \emptyset$ if $i \ne j$.

Let $d \nu_{0}:=\hat{\rho}_{0}(\widehat{X}) d \hat{\mu}$ be a probability measure on $\widehat M$ with density
\begin{equation}\label{eq:PDF0}
\hat{\rho}_{0}\left(\widehat{X}\right)=
\left\{
\begin{array}{rl}
1, \ \ &\mbox{if } \widehat{X}\in \widehat{M}_{0}\\
0, \ \ &\mbox{otherwise}\:.
\end{array}
\right.
\end{equation} 
This measure evolves under the action of $S_\alpha$ describing
  the transport of an ensemble of points initially uniformly
  distributed in $\widehat{M}_{0}$. In the following we will always
  adopt this initial setting, which mimics a $\delta$-function like
  initial condition as is common in standard diffusion theory, adapted
  to a lattice by filling a unit cell in it. If the initial condition
  were confined within $\widehat{M}_m$ with $m \ne 0$, nothing would
  change qualitatively. However, if we would fill a unit cell
  non-uniformly with particles, e.g., by choosing points close
  to the boundary of a cell, clearly we would observe very different
  dynamics. Hence, there is dependence of the outcome on the initial
  measure as is typical for IETs. Here we characterize its dynamics by
  choosing a sufficiently `nice' initial measure.

Requiring conservation of probability, the evolution $\nu_n$ of
$\nu_{0}$ at time $n$ is given by $\nu_{n}(\widehat{R})=
\nu_{0}(S^{-n}_{\alpha}\widehat{R})$ for every measurable $\widehat{R}
\subset \widehat{M}$.
Its density is
given by
\begin{equation}\label{eq:PDFN}
\hat{\rho}_{n}(\widehat{X})=
\left\{
\begin{array}{rl}
1\ \ \  & \mbox{if } \widehat{X}\in S_{\alpha}^{n} \widehat{M}_{0} \\ 
0 \ \ \  & \mbox{otherwise}\:.
\end{array}
\right.
\end{equation}
In the line above Eq.(\ref{eq:PDFN}) $S_\alpha^{-n}$ is intended in the set-theoretical sense, since
$S_\alpha^{-1} X$ is not a single point, in general.  However,
restricting to the specific initial condition given by cell
$\widehat{M}_{0}$, and to the part of $\widehat{M}$ that the points
initially in $\widehat{M}_{0}$ reach at any finite time $n$, the
preimage of a point is a single point and the inverse of the map is
defined as follows:  Consider the evolution of the initial
distribution $\widehat{N}^{(n)}=S_{\alpha}^n\widehat{M}_0,\
n=0,1,2,3,\ldots$, and define the maps
$T_n=S_{\alpha}|_{\widehat{N}^{(n)}}: \widehat{N}^{(n)} \rightarrow
\widehat{N}^{(n+1)}$ (we drop $\alpha$ from $T_n$ for the sake of
notation).  These maps are surjective. They are also
injective. Indeed, suppose $\hat{x}_1,\hat{x}_2\in\widehat{N}^{(n)}$
yields $T_n(\hat{x}_1)=T_n(\hat{x}_2)$, then
$\pi_M(T_n(\hat{x}_1))=\pi_M(T_n(\hat{x}_2))=:\xi$ and, since
$S_\alpha$ does not change the first component of any point $\hat{x}$
(i.e. $\pi_M(S(\hat{x}))=\pi_M(\hat{x}))$, we have that also
$\pi_M(\hat{x}_1)=\pi_M(\hat{x}_2)=\xi$. But in $\widehat{N}^{(n)}$
there is only one point with first component $\xi$ for any $ \xi \in
(0,1)$, thus $\hat{x}_1=\hat{x}_2$. The map that gives the evolved
distribution at time $n$ is
\begin{equation}
{\mathfrak{S}}_{\alpha,n}:=T_{n-1}\circ T_{n-2}\circ \cdots \circ T_0: \widehat{M}_0 \rightarrow \widehat{M},
\end{equation}
and the dynamics is given by the family of invertible maps $\{ {\mathfrak{S}}_{\alpha, n}\}_{n \in \N}$. Obviously ${\mathfrak{S}}_{\alpha, n}^{-1}=T_0^{-1}  \circ  \cdots \circ  T_{n-2}^{-1} \circ T_{n-1}^{-1}$. Since 
$\pi_M(T_n(\hat{x}))=\pi_M(\hat{x})$, for any $A\subset \widehat{N}_n$ we have $\lambda( \pi_M(T_n(A)))
=\lambda( \pi_M(A)) $, where $\lambda$ is the Lebesgue measure. In the same way if $A\subset \widehat{N}^{(n+1)}$, then $\lambda( \pi_M(T_n^{-1}(A)))=\lambda( \pi_M(A))$. From this it follows that if $A\subset \widehat{N}^{(0)}$, then
$\lambda(\pi_M({\mathfrak{S}}_n(A))) =\lambda( \pi_M(A))$ and if $A\subset \widehat{N}^{(n)}$ then 
$\lambda(\pi_M({\mathfrak{S}}^{-1}_{\alpha,n}(A))) =\lambda( \pi_M(A))$. In other words, maps $\{ {\mathfrak{S}}_{\alpha, n}\}_{n \in \N}$ preserve the Lebesgue measure and $\hat{\mu}$ is also invariant w.r.t. the same family of maps.

Apart from being formally precise with defining the inverse for our
model, it is interesting to conclude that $S^{-1}$ depends on the
initial condition. More physically speaking, this appears to be a
consequence of the spatial translational symmetry breaking with
respect to the different slicer positions in the different cells. The
lack of a more general definition of $S^{-1}$ implies in turn that the
slicer map is not time reversible invariant in the sense of the
existence of an involution \cite{Do99,Gasp,Kla06}, however, we don't
need the latter property for our calculations.

Consider now the sets
\begin{equation}
\widehat{R}_{j}:=S^{n}_{\alpha} \widehat{M_{0}} \cap {\widehat{M}_{j}} \equiv 
{\mathfrak{S}}_{\alpha,n} (\widehat{M_{0}}) \cap {\widehat{M}_{j}}  ,\quad j=-n,\ldots, n,
\end{equation}
which constitute the total phase space volume occupied at time $n$ in cell ${\widehat{M}_{j}}$. Their
measure
\begin{equation}
A_{j}:= \hat{\mu}( \widehat{R_{j}} )=\lambda(\pi_{M}(\widehat{R_{j}}))\delta_{\mathbb{Z}}(j)=\lambda(\pi_{M}(\widehat{R_{j}}))
\end{equation}
equals the probability $\nu_n(\widehat{M}_{j})$ of cell $j$ at time $n$: as $\hat{\mu}$ is  $\{ {\mathfrak{S}}_{\alpha, n}\}_{n \in \N}$-invariant and
${\mathfrak{S}}_{\alpha, n}$ are invertible, we have
$$
A_{j}=\hat{\mu}(\widehat{R}_{j})=\hat{\mu}( {\mathfrak{S}}_{\alpha, n}^{-1}(\widehat{R}_{j}))
=\hat{\mu}(\widehat{M_{0}}\cap {\mathfrak{S}}_{\alpha, n}^{-1} (\widehat {M_{j}}))=
\nu_{0}( {\mathfrak{S}}_{\alpha, n}^{-1}  (\widehat {M_{j}}))
=\nu_{0}( {S}_{\alpha}^{-n}  \widehat {M_{j}})=\nu_{n}({\widehat{M}_{j}}) ~,
$$ 
and $\sum_{j=-n}^{n} A_{j} = \hat{\mu}(\cup_{j=-n}^{n} {\mathfrak{S}}_{\alpha, n}^{-1}  
(\widehat{M}_{0}) \cap {\widehat{M}_{j}} )=
\widehat{\mu}({\mathfrak{S}}_{\alpha, n}^{-1}  (\widehat{M_{0}})
)=\hat{\mu}(\widehat{M_{0}}) = 1$. Indeed, ${\mathfrak{S}}_{\alpha, n}^{-1}  
(\widehat{M_{0}}) \cap {\widehat{M}_{j}} =\emptyset$ for $|j|>n$ and
$\cup_{j=-\infty}^{\infty } {\widehat{M}_{j}} = \widehat{M}$.  In
other words, the $A_{j}$'s define a probability distribution which
coincides with $\nu_{n}(\pi_{\mathbb Z}^{-1})$ and, thus, is a
marginal probability distribution of $\nu_{n}$. 
Starting from the ``microscopic'' distribution
$\nu_n$ on $\widehat{M}$, we can now introduce its coarse grained
version $\rho_{n}^{G}$ as the following measure on the integer numbers
$\mathbb{Z}$: For every time $n\in\mathbb{N}$, the coarse grained
distribution is defined by
\begin{equation}
\rho^{G}_{n}(j)=
\left\{
\begin{array}{rl}
A_j \ \ & \mbox{if } j\in \{-n,\ldots, n \},\ \\
0 \ \ \  & \mbox{\rm otherwise\quad .}
\end{array}
\right.
\end{equation}
$A_{-n}$ and $A_{n}$ are called {\em traveling areas}, $A_j$ is called
{\em sub-traveling} area if $|j|<n$.

\begin{oss}
  From the definition of $S_{\alpha}$ and the initial condition
  Eq.(\ref{eq:PDF0}), we have $A_{j}=A_{-j}$ for all $j\in
  \mathbb{Z}$. Thus, $\rho^{G}_{n}(j)$ is even,
  $\rho^{G}_{n}(j)=\rho^{G}_{n}(-j)$, and all its odd moments vanish.
\end{oss}

The coarse grained distribution will be used to describe the transport
properties of the coarse grained trajectories
$\{\pi_{\mathbb{Z}}(S^{n}_{\alpha}\widehat{X}_{0})\}_{n\in
  \mathbb{N}}\subset \mathbb{Z}$, with $\widehat{X}_{0} \in
\widehat{M}_{0} $.  This way, $\rho_{n}^{G}$ becomes the discrete
analog of the mass concentration used in ordinary and generalized
diffusion equations for systems that are continuous in time and space
\cite{MeKl00,KRS08}. Accordingly, we can define a discrete version of
the MSD as
\begin{equation}\label{eq:DMSD}
\langle \Delta \hat{X}^{2}_{n}\rangle := \sum_{j=-n}^{n} A_j j^{2}
\end{equation}
for $\rho_{n}^{G}$, where $j$ is the distance travelled by a point in
$\widehat{M}_j$ at time $n$.  Then, for $\gamma \in [0,2]$ let
\begin{equation}
T_{\alpha}(\gamma):=\lim_{n\rightarrow \infty} \frac{\langle \Delta \hat{X}^{2}_{n}\rangle }{n^{\gamma}} ~.
\end{equation}
If $T_{\alpha}(\gamma^t)\in (0,\infty)$ for $\gamma^{t}\in [0,2]$,
$\gamma^{t}$ is called the transport exponent of the slicer dynamics,
and $T_{\alpha}(\gamma^{t})$ yields the generalized diffusion
coefficient \cite{MeKl00,KRS08}.

\begin{oss}
  Due to symmetry the mean displacement $\langle \Delta
  \hat{X}_{n}\rangle := \sum_{j=-n}^{n} A_j j$ vanishes at all $n$,
  hence there is no drift in the slicer dynamics.
\end{oss}

Note that $A_{j}$ equals the width of the interval $R_{j}$, which is determined by the position of the slicers in the 
$j$-th cell, once $\alpha$ is given. Therefore, $A_{j}$ can be computed directly from Eq.(\ref{eq:SLICEGEN}). 
For the traveling areas we have
\begin{equation}\label{eq:AV}
A_n=\ell_{n-1}=\left(\frac{1}{|n|-1+2^{1/\alpha}}\right)^{\alpha} = A_{-n}
\end{equation}
while for the non vanishing sub-traveling areas we have
\begin{equation}
A_j = \ell_{\left|j\right|-1}-\ell_{\left|j\right|+1} =
\frac{1}{\left(\left|j\right|-1+2^{1/\alpha}\right)^{\alpha}}-
\frac{1}{\left(\left|j\right|+1+2^{1/\alpha}\right)^{\alpha}}~,
\end{equation}
For even $n>2$ this implies
\begin{equation}\label{eq:DENSFUNKP}
{\rho}^{G}_{n}(j)=
\left\{
\begin{array}{rl}
2(\ell_{0}-\ell_{1}) ~,       & \ \ \mbox{for }  j=0\\
\ell_{\left|2k-1\right|}-\ell_{\left|2k+1\right|} ~, & \ \ \mbox{for } |j|=2k,\, \ k=1,\ldots,\frac{n-2}{2}\\
\ell_{\left|n-1\right|} ~,           & \ \ \mbox{for } |j|=n\\
0 ~,                & \ \ \mbox{elsewhere}
\end{array}
\right.
\end{equation}
while for odd $n>3$ it implies
\begin{equation}\label{eq:DENSFUNKD}
{\rho}^{G}_{n}(j)=
\left\{
\begin{array}{rl}
\ell_{\left|2k\right|}-\ell_{\left|2k+2\right|}~, & \ \ \mbox{for }|j|=2k+1,\ k=0,\ldots,\frac{n-3}{2}\\
\ell_{\left|n-1\right|} ~,         & \ \ \mbox{for } |j|= n\\\
0 ~,                & \ \ \mbox{elsewhere}\quad .
\end{array}
\right.
\end{equation}

\begin{oss}\label{rem:pdfmsd}
  Using Eq.(\ref{eq:SLICEGEN}) in
  Eqs.(\ref{eq:DENSFUNKP},\ref{eq:DENSFUNKD}) for large $n$, one
  obtains that the tail of the distribution (large $j$) goes
  (independently of the parity of $n$) like ${\rho}^{G}_{n}(j)\sim
  {\displaystyle 2\alpha}/{\displaystyle |j|^{\alpha+1}}{\mathbb
    I}_{\{|j|<n\}}$, i.e.\ ${\rho}^{G}_{n}$ has heavy tails. Note that
  for $\alpha\in [0,2)$ these tails correspond exactly to the ones of a
  L\'evy stable distribution \cite{KRS08}. However, for $j=\pm n$
  the probability is much larger, ${\rho}^{G}_{n}(n)\sim
  {\displaystyle 1}/{ \displaystyle |n|^{\alpha}}$.  \end{oss}

We are now prepared to prove the following result: 

\begin{prop}\label{prop:msd}
  Given $\alpha\in\left(0,2\right)$ and the uniform initial
  distribution in $\widehat{M}_0$, we have
\begin{equation} \label{eq:GDIFFCO1}
T_{\alpha}(\gamma)=
\left\{
\begin{array}{rl}
+\infty               & \mbox{if }  0\leq\gamma<2-\alpha \\
\frac{4}{2-\alpha}    & \mbox{if }  \gamma = 2-\alpha    \\
0                     & \mbox{if }  2-\alpha<\gamma\leq 2 \quad ,              
\end{array}
\right. 
\end{equation}
hence the transport exponent $\gamma^{t}$ takes the value
$2-\alpha$ with $\langle\Delta \hat{X}^{2}_{n}\rangle \sim n^{2-\alpha}$. 
For $\alpha=2$ the transport regime is {\em logarithmically diffusive}, i.e.\
\begin{equation}\label{eq:logdiff}
\langle\Delta \hat{X}^{2}_{n}\rangle \sim \log n
\end{equation}
asymptotically in $n$. 
\end{prop}

\proof 
Because of the symmetry of $\rho^{G}_{n}$ let us consider only the
cells $\widehat{M}_j$ with $j \in \mathbb{N}$, so to obtain:
\begin{equation}\label{eq:AV+AQP}
T_{\alpha}(\gamma)=2 \lim_{n \to \infty} \frac{1}{n^{\gamma}}\sum_{j=0}^{n} A_j j^{2}
= 2 \lim_{n \to \infty} \frac{1}{n^{\gamma}}\left(\sum_{j=0}^{n-1} A_j j^{2} + A_n n^{2}\right).
\end{equation}
Because of Eq.(\ref{eq:AV}), the travelling area yields
\begin{equation}
\lim_{n\to \infty}\frac{n^{2}}{(n+2^{1/\alpha}-1)^{\alpha}}\cdot \frac{1}{n^{\gamma}}=
\left\{
\begin{array}{rl}
 \infty      &       \mbox{if } 0\leq\gamma<2-\alpha \\
1             &       \mbox{if } \gamma = 2-\alpha \\
0             &       \mbox{if } 2-\alpha<\gamma\leq2\quad .
\end{array}
\right.
\end{equation}
For the sub-travelling areas, by introducing $Q_{n}:=\sum_{j=0}^{n-1} A_j
j^{2}$, we will show below that
\begin{equation} \label{eq:LIMAQP}
\lim_{n \to \infty} \frac{Q_{n}}{n^{\gamma}} =
\left\{
\begin{array}{rl}
\infty \              & \ \ \mbox{if } 0\leq\gamma<2-\alpha \\
\dfrac{\alpha}{2-\alpha}  & \ \ \mbox{if } \gamma = 2 - \alpha \\
0                     & \ \ \mbox{if } 2 - \alpha < \gamma \leq 2\quad .
\end{array}
\right.
\end{equation}
{\begin{oss}\label{rem:scal} Note that the traveling and the
sub-traveling areas produce exactly the same scaling for the MSD. We
will come back to this fact in
    Section~\ref{sec:stochthy}. This result can also be obtained by
    calculating the second moment of the probability distributions of
    these two different areas directly from the expressions given in
    Remark~\ref{rem:pdfmsd} above.
\end{oss}

To prove Eq.(\ref{eq:LIMAQP}), consider that the series $Q_{n}$ assumes a
different form depending on wheter $n$ is even or odd.  If it is even
and larger than $2$ we have
\begin{equation}
Q_{n}=\sum_{j=0,j\in P_n}^{n-1}A_j j^{2} =4\sum_{j=1}^{\frac{n}{2}-1}\left[\frac{1}{\left(2j+2^{1/\alpha}-1\right)^{\alpha}}-\frac{1}{\left(2j+2^{1/\alpha}+1\right)^{\alpha}}\right]\;j^{2}\quad .
\end{equation}
This sum has a telescopic structure that allows us to rewrite it as
\begin{equation}\label{eq:REDSUM}
Q_{n}=4~\sum_{j=1}^{\frac{n}{2}-1}\frac{2j-1}{\left(2j-1+2^{1/\alpha}\right)^{\alpha}}-
\frac{\left(n-2\right)^{2}}{\left(n-1+2^{1/\alpha}\right)^{\alpha}}\quad .
\end{equation}
Let $R_{n}$ be the first term of $Q_n$.  Introducing $f(j):=\frac{2j-1}{\left(2j-1+2^{1/\alpha}\right)^{\alpha}}$ we can write
\begin{equation}
R_{n}=4~\sum_{j=1}^{\frac{n}{2}-1}\frac{2j-1}{\left(2j-1+2^{1/\alpha}\right)^{\alpha}}
= 4~\sum_{j=1}^{\frac{n}{2}-1} f(j)\quad .
\end{equation}
The derivative
\begin{equation}
f'(j)=\frac{2\left[2(1-\alpha)j+2^{1/\alpha}+\alpha-1\right]}{(2j+2^{1/\alpha}-1)^{\alpha+1}}
\end{equation}
shows that $f$ is increasing for $0<\alpha\leq1$, while for
$1<\alpha<2$, $f$ grows for $j<j(\alpha)$ and decreases for
$j>j(\alpha)$, with $j(\alpha)=(1-\alpha-2^{1/\alpha})/2(1-\alpha)$.
For $0<\alpha \le 1$ $f$ is strictly increasing, hence
\begin{equation}\label{eq:DIS1}
\int_{0}^{\frac{n}{2}-1}f(x)dx\leq\sum_{j=1}^{\frac{n}{2}-1}f(j)\leq\int_{1}^{\frac{n}{2}}f(x)dx\quad .
\end{equation}
We have to distinguish two cases, $\alpha<1$ and $\alpha=1$. In the
first case we have
\begin{eqnarray}
\int_{0}^{\frac{n}{2}-1}f(x)dx &=& \frac{1}{2}\left[\frac{(n-3+2^{1/\alpha})^{2-\alpha}}{2-\alpha}-2^{1/\alpha}\cdot\frac{(n-3+2^{1/\alpha})^{1-\alpha}}{1-\alpha} \right.+ \nonumber \\
&+& \left. \left(2^{1/\alpha}-1\right)^{1-\alpha}\cdot\frac{2^{1/\alpha}-\alpha+1}{(2-\alpha
)(1-\alpha)}\right]
\end{eqnarray}
and
\begin{eqnarray}
\int_{1}^{\frac{n}{2}}f(x)dx &=& \frac{1}{2}\left[\frac{(n-1+2^{1/\alpha})^{2-\alpha}}{2-\alpha}-2^{1/\alpha}\cdot\frac{(n-1+2^{1/\alpha})^{1-\alpha}}{1-\alpha} \right.+ \nonumber \\
&+& \left.(2^{1/\alpha}+1)^{1-\alpha}\cdot\frac{(2^{1/\alpha}+\alpha-1)}{(2-\alpha)(1-\alpha)}\right]\quad ,
\end{eqnarray}
therefore taking the $n\rightarrow\infty$ limit we have
\begin{equation}\label{eq:LIMINT1}
\lim_{n \to \infty} \frac{1}{n^{\gamma}}\int_{0}^{\frac{n}{2}-1}f(x)dx=
\left\{
\begin{array}{rl}
\infty                   &       \mbox{if } 0\leq\gamma<2-\alpha \\
\frac{1}{2(2-\alpha)}    &       \mbox{if } \gamma =2-\alpha \\
0                        &       \mbox{if } 2-\alpha<\gamma\leq2
\end{array}
\right.
\end{equation}
and
\begin{equation}\label{eq:LIMINT2}
\lim_{n \to \infty} \frac{1}{n^{\gamma}}\int_{1}^{\frac{n}{2}}f(x)dx=
\left\{
\begin{array}{rl}
\infty                   &       \mbox{if } 0\leq\gamma<2-\alpha \\
\frac{1}{2(2-\alpha)}    &       \mbox{if } \gamma =2-\alpha \\
0                        &       \mbox{if } 2-\alpha<\gamma\leq2\quad .
\end{array}
\right.
\end{equation}
For $\alpha=1$ the two integrals differ, but the bounding 
limits coincide. Therefore one obtains
\begin{equation} \label{sopra}
\lim_{n \to \infty} \frac{R_{n}}{n^{\gamma}}=
\lim_{n \to \infty} \frac{4}{n^{\gamma}} \sum_{j=1}^{\frac{n}{2}-1}f(j)=
\left\{
\begin{array}{rl}
\infty                   &       \mbox{if } 0\leq\gamma<2-\alpha \\
\dfrac{2}{2-\alpha}    &       \mbox{if } \gamma =2-\alpha \\
0                        &       \mbox{if } 2-\alpha<\gamma\leq2\quad .
\end{array}
\right.
\end{equation}

\noindent
If $1<\alpha< 2$, $f$ decreases for $j>j(\alpha)$. Hence, introducing
$\bar{j}_{\alpha}=\left\lfloor j(\alpha)\right\rfloor$, where $\left\lfloor x\right\rfloor$ 
is the integer part of $x$, $R_n$ can be expressed as: 
\begin{equation}\label{eq:REDSUM2}
R_{n}=4\left(\sum_{j=1}^{\bar{j}_{\alpha}}f(j)+\sum_{j=\bar{j}_{\alpha}+1}^{\frac{n}{2}-1}f(j)\right)\quad .
\end{equation}
Dividing by $n^\gamma$ and taking the $n \to \infty$ limit, the first
term vanishes for all $\gamma>0$ while the second term can be treated
as above to obtain the same as Eq.(\ref{sopra}).  Recalling
Eq.(\ref{eq:REDSUM}), this eventually implies
Eq.(\ref{eq:LIMAQP}). For odd $n$, one proceeds similarly.

In summary, the MSD grows like $\langle \Delta \hat{X}^{2}_{n}\rangle
\sim n^{2-\alpha}$ for $\alpha\in (0,2)$, and the trivial slicer map
$S_{\alpha}$ enjoys all power law regimes of normal and anomalous
diffusion as $\alpha$ varies in $(0,2)$.

\begin{oss}\label{rem:ball} From Eq.(\ref{eq:SLICEGEN}) it
    follows trivially that for $\alpha \to 0$ we have $\ell_j=1/2,\:
    \forall j\in \mathbb{Z}$. This means that everywhere on the slicer
    lattice half unit intervals are mapped onto half unit intervals in
    neighbouring cells in the same direction of the previous jump
    generating purely ballistic motion. Consequently for $\alpha=0$
    the MSD grows like $\langle \Delta \hat{X}^{2}_{n}\rangle \sim
    n^2\:(n\to\infty)$. 
\end{oss}

Repeating the previous reasonings by computing the correspondingly
different integrals for $\alpha=2$ for a uniform initial distribution
in $\widehat{M}_0$ we find that
\begin{equation} \label{eq:GDIFFCO2}
T_{2}(\gamma)=
\left\{
\begin{array}{rl}
+\infty               & \mbox{if }\ \  \gamma=0 \\
0                     & \mbox{if } \ \  0<\gamma\leq 2\quad .
\end{array}
\right. 
\end{equation}
The upper and lower bounds of the integrals corresponding to
  Eq.(\ref{eq:DIS1}) feature leading logarithmic terms, which yields
  Eq.(\ref{eq:logdiff}).
\endproof

\begin{oss}\label{rem:loc} By an analogous calculation, or alternatively by looking at the second
moment of the probability distributions, cf.\ Remark \ref{rem:scal} above, one can see that for $\alpha>2$
\begin{equation}
\langle\Delta \hat{X}^{2}_{n}\rangle \to const.\quad (n\to\infty)\quad .
\end{equation}
That is, in terms of the MSD localisation sets in, although from the
definition of the slicer map it is intuitively not clear why this
should happen. 
\end{oss}


These results can now be generalised by calculating the asymptotic
behavior of the higher moments $\Delta \hat{X}^{p}_{n}$ of
$\rho^{G}_{n}$,
\begin{equation} \label{eq:homom}
\langle  \Delta \hat{X}^{p}_{n} \rangle = \sum_{j=-n}^{n} A_{j}j^{p}\quad .
\end{equation}
\begin{thm}\label{thm:MOM}
For $\alpha\in\left(0,2\right]$ the moments $\langle \Delta \hat{X}^{p}_{n}\rangle$ with $p>2$ even and initial 
condition uniform in $\widehat{M}_0$ have the asymptotic behavior
\begin{equation}
\langle \Delta \hat{X}^{p}_{n}\rangle \sim n^{p-\alpha} \label{eq:thmmsd}
\end{equation}
while the odd moments ($p=1,3,...$)  vanish.
\end{thm}
\proof
We want to compute the limit
\begin{equation}
L(\alpha,p):=\lim_{n\to\infty}\frac 1 {n^{\gamma}} {\langle \Delta \hat{X}^{p}_{n}\rangle}=
\lim_{n\to\infty}\ \frac 1 {n^{\gamma}}\sum_{j=-n}^{n} A_j j^{p}\quad .
\end{equation}
As observed in Remark 2, the symmetry of $\rho^{G}_{n}$ implies that
the sums with odd $p$ to vanish.  For the even moments it suffices to
consider the positive $j$'s,
\begin{equation}\label{eq:AV+AQP-MOM}
L(\alpha,p) = \lim_{n \to \infty} \frac{2}{n^{\gamma}}\left(\sum_{j=0}^{n-1} A_j j^{p} + A_n  n^{p}\right)\quad.
\end{equation}
We now prove that
\begin{equation} \label{eq:LIMAQP-MOM}
\lim_{n \to \infty} \frac{1}{n^{\gamma}}\sum_{j=0}^{n-1}A_j j^{p}=
\left\{
\begin{array}{rl}
\infty \                  & \ \ \mbox{if } 0\leq\gamma<p-\alpha \\
\dfrac{\alpha}{p-\alpha}>0 & \ \ \mbox{if } \gamma = p - \alpha \\
0                         & \ \ \mbox{if } \gamma > p - \alpha \quad .
\end{array}
\right.
\end{equation}
In order to do so, for even $n$ let us introduce
\begin{equation}\label{eq:SERIEMOM}
\mathcal{P}_{n}:=\sum_{j=0,j\in P_n}^{n-1}A_j j^{p} =
2^{p}\sum_{j=1}^{\frac{n}{2}-1}
\left[\frac{1}{\left(2j+2^{1/\alpha}-1\right)^{\alpha}}-\frac{1}{\left(2j+2^{1/\alpha}+1\right)^{\alpha}}\right]\;j^{p}\quad .
\end{equation}
A simple induction procedure leads to
\begin{equation}\label{eq:REDSUM-MOM}
\mathcal{P}_{n}=2^{p}\cdot\sum_{j=0}^{\frac{n}{2}-2}\frac{(j+1)^{p}-j^{p}}{\left(2j+1+2^{1/\alpha}\right)^{\alpha}}-
\frac{\left(n-2\right)^{p}}{\left(n-1+2^{1/\alpha}\right)^{\alpha}} = R_{n}-
\frac{\left(n-2\right)^{p}}{\left(n-1+2^{1/\alpha}\right)^{\alpha}}~,
\end{equation}
which defines $R_{n}$ in terms of addends of the form
\begin{equation}
f(j)=\frac{(j+1)^{p}-j^{p}}{(2j+1+2^{1/\alpha})^{\alpha}}=
\sum_{k=1}^{p}\binom{p}{k}\frac{j^{p-k}}{(2j+1+2^{1/\alpha})^{\alpha}}
\end{equation}
with derivatives given by
\begin{equation}
f'(j)=
\sum_{k=1}^{p}\binom{p}{k}\frac{\left[2(p-k-\alpha)j+(p-k)(1+2^{1/\alpha})\right]}{(2j+1+2^{1/\alpha})^{\alpha+1}}j^{p-k-1}
= \sum_{k=1}^{p}f_{k}(j)\quad ,
\end{equation}
where
\begin{equation}
f_{k}(j)=\binom{p}{k}\frac{\left[2(p-k-\alpha)j+(p-k)(1+2^{1/\alpha})\right]}{(2j+1+2^{1/\alpha})^{\alpha+1}}j^{p-k-1}~.
\end{equation}
For $0<\alpha\leq 1$ and all $j>0$ we have
$f_{k}(j)>0$ for $k=1,\ldots,p-1$, while $f_{p}(j)<0$. Because $| f_p(j) | < f_1(j)$,
$f'$ is positive and $f$ increases for all $j>0$.
For $1<\alpha\leq 2$ and $p=3$, one has $f(j)=(3 j^2 + 3 j +1)/(2 j + 1 + 2^{1/\alpha})^\alpha$,
which is increasing for $j>0$, while for $p \ge 4$ one obtains 
$f_{k}(j)>0$ for $k=1,\ldots,p-2$, and $f_{p-1}(j), f_{p}(j)<0$. Because
$| f_{p-1}(j) + f_p(j) | < f_1(j) + f_2(j)$, $f(j)$ is increasing for $j>0$,
even for $1<\alpha\leq 2$.

Therefore our sum is bounded from above and below by
\begin{equation}\label{eq:DIS1-MOM}
\int_{0}^{\frac{n}{2}-2}f(x)dx < \sum_{j=0}^{\frac{n}{2}-2}f(j) < \int_{1}^{\frac{n}{2}-1}f(x)dx
\end{equation}
for all  $\alpha \in (0,2]$.
Taking the limit as done previously, we eventually obtain
\begin{equation}
\lim_{n\to\infty}\frac{R_{n}}{n^{\gamma}}=
\left\{
\begin{array}{rl}
\infty                   &       \mbox{if } 0\leq\gamma<p-\alpha \\
\dfrac{p}{2^{p}(p-\alpha)}  &       \mbox{if } \gamma =p-\alpha \\
0                        &       \mbox{if } \gamma>p-\alpha\quad ,
\end{array}
\right. \quad
\lim_{n \to \infty} \frac{\mathcal{P}_{n}}{n^{\gamma}}  =
\left\{
\begin{array}{rl}
\infty                  &     \mbox{if } 0\leq\gamma<p-\alpha \\
\dfrac{\alpha}{p-\alpha} &     \mbox{if } \gamma = p-\alpha    \\
0                       &     \mbox{if } \gamma>p-\alpha
\end{array}
\right.
\end{equation}
for all $\alpha\in (0,2]$. If $n$ is odd one proceeds similarly to
obtain the same result.

For the travelling area Eq.(\ref{eq:AV}) we have
\begin{equation}
\lim_{n\to \infty} A_n n^{p}=\lim_{n\to \infty}\frac{n^{p}}{(n+2^{1/\alpha}-1)^{\alpha}}\cdot \frac{1}{n^{\gamma}}=
\left\{
\begin{array}{rl}
\infty      &       \mbox{if } 0\leq\gamma<p-\alpha \\
1             &       \mbox{if } \gamma = p-\alpha \\
0             &       \mbox{if } \gamma > p - \alpha \quad .
\end{array}
\right.
\end{equation}
Hence, the same scaling for traveling and sub-traveling regions as
pointed out for the second moment in Remark~\ref{rem:scal} holds for
all higher moments. We thus conclude that
\begin{equation}
L(\alpha,p) = 
\left\{
\begin{array}{rl}
\infty                      &      \mbox{if } 0\leq\gamma<p-\alpha \\
\dfrac{p}{p-\alpha}   &      \mbox{if } \gamma = p-\alpha    \\
0                           &      \mbox{if } \gamma > p - \alpha
\end{array}
\right.
\end{equation}
so that the large $n$ behavior of the even moments is given by
$\langle \Delta \hat{X}^{p}_{n}\rangle \sim n^{p-\alpha}$.
\endproof

\subsection{Example}
In this subsection we illustrate the diffusive transport generated by
the slicer map $S_{\alpha}$ for a representative value of
$\alpha$. For this purpose we plot the probability distribution
  using our exact analytical results and compare it to an asymptotic
  approximation. We then draw cross-links to diffusive transport in a
  polygonal channel.

As an example, let us consider the case $\alpha=1/3$. Why we choose
this particular values is explained further below. Here we have
$\ell_j(1/3)={1}/{\left(\left|j\right|+8\right)^{1/3}}$, and the
asymptotic behavior of the MSD is given by $\langle \Delta
\widehat{X}^{2}_{n}\rangle \sim n^{{5}/{3}}$, cf.\
Proposition~\ref{prop:msd}.  This means that $S_{1/3}$ is
superdiffusive with $\gamma^t=5/3$ and generalized diffusion
coefficient $T_{1/3} ={12}/{5}$. From Theorem \ref{thm:MOM} the
moments of $S_{1/3}$ higher than the second have the behavior
\begin{equation}
\langle \Delta \widehat{X}^{p}_{n}\rangle \sim n^{p-{1}/{3}}\:.
\end{equation}
The coarse grained distribution for even $n$ reads, see
Eqs.(\ref{eq:DENSFUNKP},\ref{eq:DENSFUNKD}),
\begin{equation}
\rho^{G}_{n}\left(m\right)=
\left\{
\begin{array}{rl}
\frac{1}{2}-\frac{1}{\sqrt[3]{9}},       & \ \ \mbox{for } m=0\\
\frac{1}{\sqrt[3]{2k+7}}-\frac{1}{\sqrt[3]{2k+9}}, & \ \ \mbox{for } m=2k,\ k=2,\ldots,\frac{n-2}{2}\\
\frac{1}{\sqrt[3]{n+7}}        & \ \ \mbox{for } m=n\\
0,                & \ \ \mbox{otherwise}\: ,
\end{array}
\right. \label{eq:ex2e}
\end{equation}
while for odd $n$ we have
\begin{equation}
\rho^{G}_{n}\left(m\right)=
\left\{
\begin{array}{rl}
\frac{1}{\sqrt[3]{2k+8}}-\frac{1}{\sqrt[3]{2k+10}}, & \ \ \mbox{for } m=2k+1,\ k=2,\ldots,\frac{n-3}{2},\\
\frac{1}{\sqrt[3]{n+7}}       & \ \ \mbox{for } m=n\\
0,                & \ \ \mbox{otherwise}\:.
\end{array}
\right. \label{eq:ex2o}
\end{equation}
Figure \ref{fig:DENS7CONF} shows the marginal probability
distribution function $\rho^{G}_{n}(m)$ at fixed even $n$ for $m>0$,
including the
last value $\rho^{G}_{n}(n)$ which is much larger than the values for
$m$ close to $n$. The negative branch of the distribution can be
recovered by symmetry.

Because asymptotically $\rho^{G}_{n}$ goes like
\begin{equation}\label{eq:ROX}
\rho^{\alpha}_{n}(m)=
\left \{
\begin{array}{rr}
\frac{\displaystyle C_{\alpha}}{\displaystyle ( m+2^{1/\alpha})^{\alpha+1}}~, & m<n,\\
0 ~, & m>n,
\end{array}
\right.
\end{equation} 
where $C_{\alpha}$ is a normalization constant,
Fig.~\ref{fig:DENS7CONF} compares the numerical values of
$\rho^{G}_{n}$ with our asymptotic result for $\rho^{1/3}_{n}(m)$ and
$C_{1/3}=1$. Apart from the peak at $\rho^{G}_{n}(n)$ due to the
traveling area, which is covered by Eqs.(\ref{eq:ex2e},\ref{eq:ex2o}),
the asymptotic behavior of both results is clearly the same.  Note
that the spike at $m=n$ is analogous to the one found in
\cite{BFK00,BCV10}.

The choice of $\alpha=1/3$ is motivated by results on diffusion
in the sawtooth polygonal channel studied in \cite{JeRo06}. The
channel geometry is analogous to the one shown in
Fig.~\ref{fig:poly_chann} except that there are no flat wall
sections between any two triangles. The angle between one side of a
triangle and the wall base line has been chosen to
$\pi/4$. Simulations for this particular case yielded a transport
exponent of about $\gamma_t=5/3$, cf.\ Table 1 in \cite{JeRo06}. This
result is surprising, as naively one would have expected irrational
polygons to generate transport close to diffusion, and rational
polygons to exhibit transport close to ballistic. From this
perspective the case of $\pi/4$ angles should have been perfectly
ballistic, while it turned out to be substantially slower than all
other irrrational cases with parallel walls reported in \cite{JeRo06}.
This suggests that the main mechanism generating diffusion in this
channel may have less to do with whether respective polygons are
rational or irrational but rather how precisely they slice a beam of
diffusing particles as modelled by our slicer
dynamics. Fig.~\ref{fig:slicer} indicates that in this case
diffusion may be slowed down due to an incresing fraction of orbits
being localized by not contributing to diffusion. This reminds of
similar findings for polygonal billiard channels presented in
\cite{SaLa06}.

We remark that results completely analogous to
Fig.~\ref{fig:DENS7CONF} are obtained for any other value of
$0<\alpha<2$. This implies that for $\alpha=1$ our system generates a
very strange type of normal diffusion with a non-Gaussian probability
distribution.  For $1<\alpha<2$ it is furthermore surprising that
ballistic peaks representing traveling regions are present while the
model as a whole exhibits subdiffusion. We are not aware of results in
the literature where subdiffusive dynamics with coexisting traveling
regions has been observed.

\begin{figure}
\centering
\includegraphics[scale=0.5]{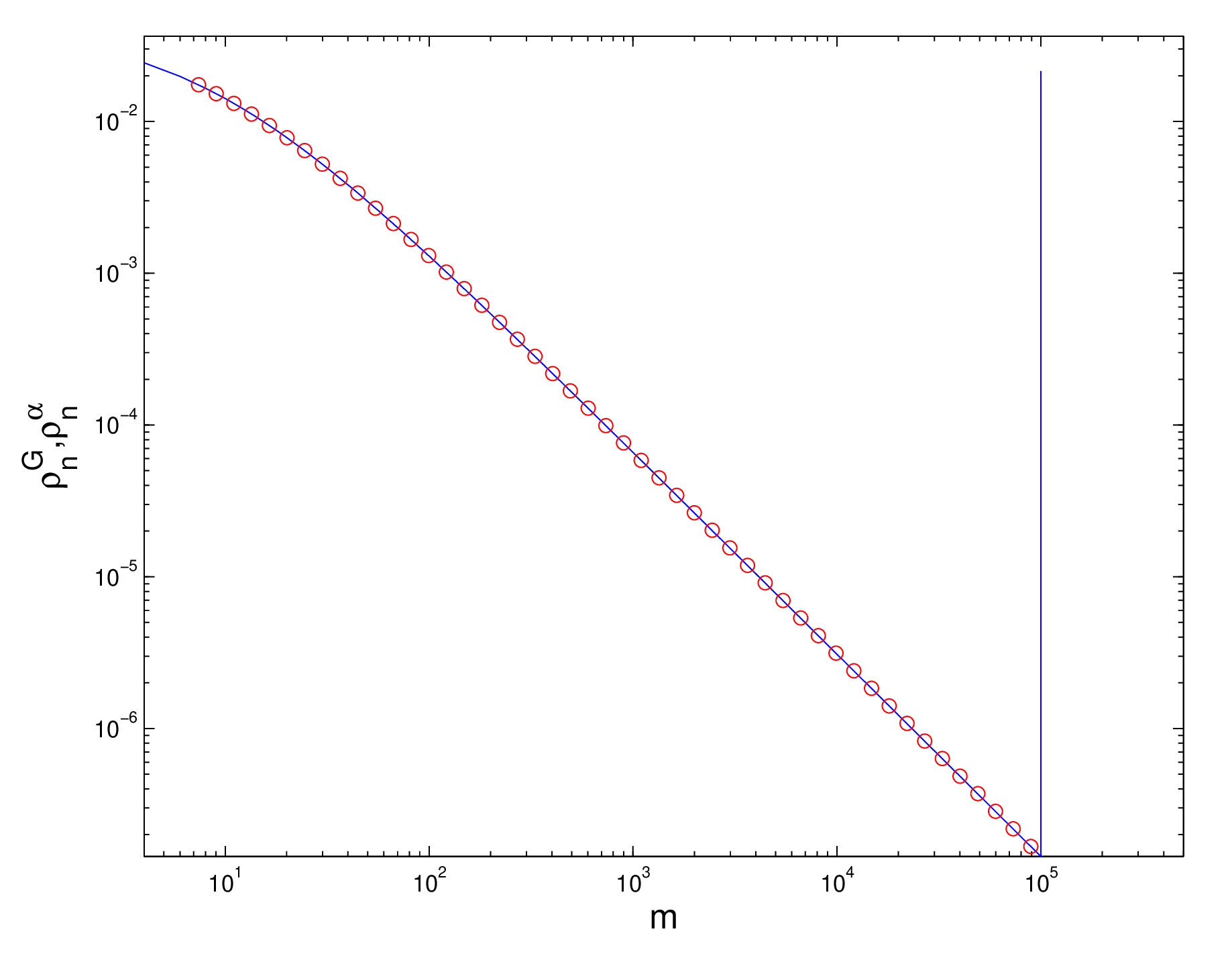}
\caption{Log-log pot of the marginal probability distribution for the
  slicer map $S_{1/3}$ as a function of the position of the $m$-th
  cell at fixed time $n=10^{5}$.  $\rho_{n}^{G}(m)$ denotes the
    coarse grained distribution obtained from our exact analytical
    results Eqs.(\ref{eq:ex2e},\ref{eq:ex2o}) (continuous line). It is
    compared with the asymptotic analytical approximation
    $\rho_{n}^{\alpha}(m)$, Eq.(\ref{eq:ROX}) (dotted line). Apart
  from the values at $m=n$, where $\rho_{n}^{G}$ has a spike due to
  the travelling area, the asymptotic functional forms coincide.}
\label{fig:DENS7CONF}
\end{figure}

\section{A simple stochastic model of slicer
  diffusion?}\label{sec:stochthy}

Since deterministic dynamical systems often generate a type of
randomness, it is frequently attempted to match their dynamics to
simple stochastic processes for understanding their transport
properties \cite{Do99,Gasp,Kla06,RoMM07,CFLV08}. However, as we
pointed out in the introduction, for diffusion in polygonal billiards
such a stochastic modeling turned out to be surprisingly non-trivial
\cite{Zas02,LWWZ05,Kla06,Dett14}. Motivated by this line of research,
in this section we relate the slicer diffusion to known stochastic
models of anomalous diffusion.

We first summarize our main results for diffusive transport generated
by our non-chaotic slicer map under variation of the control parameter
$0\le\alpha\le2$. In the limit of $n\to\infty$ we have:

\begin{enumerate}
\item $\alpha=0$: ballistic motion with MSD $\langle \Delta
  \hat{X}^2_{n}\rangle \sim n^2$
\item $0<\alpha<1$: superdiffusion with MSD 
$\langle \Delta \hat{X}^2_{n}\rangle \sim n^{2-\alpha}$
\item $\alpha=1$: normal diffusion with linear MSD $\langle \Delta
  \hat{X}^2_{n}\rangle \sim n$
\item $1<\alpha<2$: subdiffusion with MSD $\langle \Delta
  \hat{X}^2_{n}\rangle \sim n^{2-\alpha}$
\item $\alpha=2$: logarithmic subdiffusion with MSD $\langle \Delta
  \hat{X}^2_{n}\rangle \sim \log n$
\item $\alpha>2$: localisation in the MSD with $\langle \Delta
  \hat{X}^2_{n}\rangle \sim const.$
\end{enumerate}

Additionally, Theorem~\ref{thm:MOM} gives information about the
asymptotic behavior of all higher order even moments scaling as
$\langle \Delta \hat{X}^{p}_{n}\rangle \sim n^{p-\alpha}$ in the long
time limit for $p>2$ and $0<\alpha\le2$.

As recently highlighted in Refs.~\cite{KHL10,HKL14}, there do not seem
to exist too many stochastic models exhibiting a transition from
subdiffusion over normal diffusion to superdiffusion under parameter
variation: We are aware of a specific continuous time random walk
(CTRW) model \cite{ZuKl95}, (generalized) elephant walks
\cite{KHL10,HKL14}, and generalized Langevin equations (gLe) including
fractional Brownian motion \cite{PWM96,TCK10,JM10}. For these models one can
easily check that there is no simple matching between their diffusive
properties and the above scenario representing slicer diffusion. That
is, the scaling of the MSD with parameters by switching between all
diffusive regimes is generically different from the slicer dynamics,
and/or more than one control parameter is needed to change the
diffusive properties. However, for both the CTRW and the gLe there is
a partial matching to the slicer diffusion in specific diffusive
regimes to which we come back below. Meaning so far we are not aware
of any stochastic model that (asymptotically) reproduces the slicer
diffusion by exhibiting all the different diffusive regimes listed
above under single parameter variation.

For all the stochastic models just mentioned the dynamics is generated
by temporal randomness, that is, random variables are drawn in time
from given probability distributions. A second fundamental class of
stochastic models is defined by spatial randomness of the positions of
scatterers with a point particle moving between them. An important
example is the one-dimensional stochastic L\'evy Lorentz gas (LLg)
\cite{BFK00,BF99,BCV10}: Here a point particle moves ballistically
between static point scatterers arranged on a line from which it is
either transmitted or reflected with probability $1/2$. The
distance $r$ between two consecutive scatterers is a random variable drawn
independently and identically distributed from a L\'evy distribution, 
\begin{equation}
\lambda(r)\equiv\beta r_{0}^{\beta}\frac{1}{r^{\beta+1}}, \ \ r\in\left[r_{0},+\infty\right)\:,
\label{lamb}
\end{equation}
where $\beta>0$ and $r_{0}$ is a cutoff fixing the characteristic
length scale of the system.  The LLg shares a basic similarity with
the slicer map in that its scatterers are positioned according to the
same asymptotic functional form as the slicers, see
Eq.(\ref{eq:SLICEGEN}). On the other hand, the slicer positions are
deterministic while the LLg scatterers are distributed randomly. In
more detail, the slicers amount to transition probabilities for
nearest neighbour jumps on a periodic lattice that decrease by a power
law while in the LLg the jumps follow a power law distribution with
trivial transition probabilites at random scatterers. Finally, the
slicer dynamics is discrete in time while the LLg is a time-continuous
system.

 From these facts it follows that there exists an intricate
  dependence on initial conditions in the LLg that is not present that
  way in the slicer map: The LLg diffusive properties depend on
  whether a walker starts anywhere on the line, which means typically
  between two scatterers, called equilibrium initial condition, or
  exactly at a scatterer, called nonequilibrium initial condition
  \cite{BFK00,BCV10}. In \cite{BFK00} bounds for the MSD have been
  calculated in both cases. Interestingly, the lower bound for
  equilibrium initial conditions was obtained from CTRW theory on
  which we will elaborate further below.

  The results for nonequilibrium conditions have been improved in
  \cite{BCV10} based on simplifying assumptions by which asymptotic
  results for the position probability distribution of the moving
  particle could be calculated. It was shown that the LLg only
  displays superdiffusion, which as in the slicer map is governed by
  traveling (in \cite{BCV10} called subleading) and sub-traveling (in
  \cite{BCV10} called leading) contributions. However, while in the
  slicer map both these contributions scale in the same way for the
  MSD, cf.\ Remark~\ref{rem:scal}, for the LLg they yield different
  scaling laws depending on the order of the moment $p$ and the
  control parameter $\beta$. These different regimes are deeply rooted
  in the different physics of the system featuring an intricate
  interplay between different length scales. As a consequence, the LLg
  MSD displays three different regimes with an exponent determined by
  three different functions of $\beta$, cf.\ Eq.(3) in
  \cite{BCV10}. All higher order moments could also be calculated for
  the LLg. Very interestingly, Eq.(13) in \cite{BCV10} and all even
  moments of the slicer map, cf.\ Theorem~\ref{thm:MOM}, exactly
  coincide by a piecewise transformation between $\beta\in(0,3/2]$ and
  $\alpha\in(0,1]$. In other words, for every $\beta$ in \cite{BCV10}
  it suffices to fix the slicer's parameter $\alpha$ so that one of
  the moments ({\em e.g.} the second) matches, to match all other
  moments as well \footnote{The slicer odd moments vanish because of
    the symmetry of the map considered here. If the moments are
    computed only over the positive half of the chain, also the odd
    moments can be identified with those of \cite{BCV10}.}. Then the
  asymptotic form of the moments $\langle r^{p}(t)\rangle$ for all
  $p>0$ is given by
\begin{equation}\label{eq:MOMBUR}
\langle r^{p}(t)\rangle\sim
\left\{
\begin{array}{rl}
t^{\frac{p}{1+\beta}} ~,                        & \ \mbox{if } \beta<1,\ p<\beta \\
t^{\frac{p(1+\beta)-\beta^{2}}{1+\beta}} ~,     & \ \mbox{if } \beta<1,\ p>\beta\\
t^{\frac{p}{2}} ~,                              & \ \mbox{if } \beta>1,\ p<2\beta-1\\
t^{\frac{1}{2}+p-\beta} ~,                      & \ \mbox{if } \beta>1,\ p>2\beta-1  
\end{array}
\right.\:.
\end{equation}
Surprisingly they can be matched with the slicer moments in
Eq.(\ref{eq:thmmsd}) by taking
\begin{equation}
\alpha =
\left\{
\begin{array}{rl}
{\beta^{2}}/{(1+\beta)} & \ \ \mbox{if } 0<\beta\leq1 \\
\beta - {1}/{2} &\ \ \mbox{if } 1<\beta\leq\frac{3}{2}\\
1 & \ \ \mbox{if } \beta>\frac{3}{2}
\end{array}
\right.\:.\label{eq:llgpdf}
\end{equation}
This means that by using the above transformation both processes are
asymptotically indistinguishable from the viewpoint of these moments,
meaning the slicer map generates a kind of LLg-type walk in the
superdiffusive regime if the available information on the system (the
observables) include the moments only. On the other hand, the
transformation is piecewise which reflects the different functional
forms for the exponent of the moments in the LLg while for the slicer
map only one such functional form exists. Indeed, it is well known
that the moments carry only partial information on the properties of
(anomalous) transport phenomena, and that knowledge of correlations is
necessary to distinguish one class of stochastic processes from
another \cite{KRS08,Sokolov}.

Another interesting fact within this context is that for the traveling
region alone (called ballistic contribution in \cite{BFK00}) the MSD
of the LLg scales as $\sim t^{2-\beta}$ in continuous time $t$, as was
shown in \cite{BFK00}. Formally, this result matches exactly to the
MSD of the slicer map of $\sim n^{2-\alpha}$ as calculated in
Proposition~\ref{prop:msd}. Note also that the slicer positions
generate a probability distribution for the sub-traveling region of
${\rho}^{G}_{n}(j)\sim |j|^{-\alpha-1}$, see Remark~\ref{eq:SLICEGEN},
which matches to $\lambda(r)$ of the LLg Eq.~(\ref{lamb}).

We now comment on similarities and differences of the slicer diffusion
with CTRW theory. For equilibrium initial conditions in the LLg it was
shown that for $1<\beta<2$ the MSD is bounded from below by $\sim
t^{3-\beta}$ \cite{BFK00}. However, this is exactly the result for a
L\'evy walk modeled by CTRW theory \cite{ZuKl95}. Even more, results
for all higher L\'evy walk CTRW moments were recently calculated to
$\sim t^{p+1-\beta}$ for $p>\beta$, see Eq.(18) in \cite{RSHB14}. With
$\beta=1+\alpha$ the slicer superdiffusion thus formally (also)
matches to L\'evy walk diffusion defined by CTRW theory.  On the other
hand, from a conceptual point of view a CTRW is constructed very
differently from both the LLg and the slicer map dynamics. Hence, it
is not very clear why a CTRW mechanism should apply in both these
cases. Another remark is that for the superdiffusive regime of the
slicer diffusion of $0\le\alpha<1$ the frozen L\'evy distribution
according to which the transition probabilities have been defined does
not belong to the parameter regime for which such distributions are
stochastically stable in the sense of a generalized central limit
theorem \cite{MeKl00,KRS08}, which holds only for $1<\alpha\le3$. This
points again at a crucial difference between slicer dynamics and CTRW
theory, where for the latter the resulting probability distributions
are stochastically stable. Our discussion suggests that there is a
more intricate interplay between the L\'evy potential we {want to
  mimic, the dynamics we use to obtain it, and the power law
  distributions we generate from it.

We conclude this section by a remark on a curious similarity between
the slicer diffusion in the subdiffusive regime and Gaussian
stochastic processes. It was shown in Refs.~\cite{PWM96,TCK10,ChKl12} that
for a gLe with power law memory kernels for friction and/or noise the
MSD exhibits a transition from a constant over $\sim\log n$ to
subdiffusion in the long time limit. Especially, for an overdamped gLe
with Gaussian noise governed by a power law anti-persistent memory
kernel $\sim -t^{-\gamma}$ the MSD was calculated to $\sim
t^{2-\gamma}$ for $1<\gamma<2$, $\sim \log t$ for $\gamma =2$ and
$\sim const. $ for $\gamma>2$ \cite{ChKl12}. Formally these results
match exactly to the slicer MSD for $1<\alpha$, cf.\
Proposition~\ref{prop:msd}. However, this overdamped gLe does not
exihibit any superdiffusion. And as the probability distributions
generated by such a gLe are strictly Gaussian in the long time limit
there is, again, a clear conceptual mismatch to the slicer dynamics:
In the gLe the subdiffusion is generated from power law memory in the
random noise (by calculating the MSD via the Taylor-Green-Kubo
formula \cite{PWM96,TCK10,ChKl12}) while for the slicer map the anomalous
dynamics was calculated from non-Gaussian probability
distributions. We remark that so far nothing is known about the
correlation decay in the slicer dynamics; for the LLg it is very
complicated \cite{BF99}. Hence, while formally there might be a
similar mechanism in gLe and slicer dynamics for generating
subdiffusion, again, conceptually these dynamics are very different.

To summarize this section, to our knowledge currently there is
no stochastic model that fully reproduces the slicer diffusion. The
superdiffusive slicer regime formally matches to diffusion known from
L\'evy walks, as reproduced both by the LLg and CTRW theory, although
in detail the parameter dependences for the MSD generated by both
models are different. The subdiffusive regime formally matches to what
is generated by a gLe with power law correlated Gaussian noise.
Conceptually all these stochastic models are very different from the
slicer model, hence any similarity is purely formal and not supported
from first principles. Based on this analysis it is tempting to
conclude that one may need a correlated CTRW model to stochastically
reproduce the slicer diffusion, or perhaps alternatively a simple
L\'evy Markov chain model.

The reason why we elaborated so explicitly on a possible stochastic
modeling of the slicer dynamics is that our analysis might help to
understand the reason for the controversy of how to stochastically
model diffusion in polygonal billiards
\cite{Zas02,LWWZ05,Kla06,Dett14}. With the slicer map we are trying to
capture a basic mechanism generating diffusion in these
systems. However, as we showed above, the slicer diffusion seems to
share features with generically completely different stochastic models
depending on parameter variation. This might help to explain why
different groups of researchers came to contradicting conclusions for
modeling diffusion in polygonal billiards (not from first principles)
by applying different types of stochastic processes.

\section{Concluding remarks}

In search of mathematically tractable deterministic models of normal
and anomalous diffusion, which may shed light on the minimal
mechanisms generating different transport regimes in non-chaotic
systems, we have introduced a new model which we called the slicer
map. This map mimics in a one dimensional space main features that
distinguish periodic polygonal billiards from other models of
transport, namely the complete absence of randomness and of positive
Lyapunov exponents, and a sequence of splittings of a beam of
particles due to collisions at singularities of the billiard walls. As
observed in Refs.~\cite{ArGuRe00,ARV03,SchSt06,JeRo06,SaLa06,JBR08},
in these cases the geometry determines the transport law, differently
from standard hydrodynamics in which the geometry only yields the
boundary conditions. Therefore the rule according to which the
polygonal scatterers are distributed in space plays a crucial
role. Here we have investigated the case of a specific deterministic
rule modeling diffusion in polygonal billiards.

In our one-dimensional slicer model $S_{\alpha}$ the effect of the
billiard geometry, which ``slices'' beams of particles ever more
finely further and further away from the origin, is produced by the
rate at which the size of the slices decreases with the position,
i.e.\ by the value of a single control paraemeter $\alpha$.  For
instance, $\alpha={1}/{3}$ yields for the MSD and for the even higher
order moments $ \langle \Delta \widehat{X}^{2}(n) \rangle\ \sim
n^{\frac{5}{3}} \quad \mbox{and} \quad \langle \Delta
\widehat{X}^{p}(n) \rangle \sim n^{\frac{3p-1}{3}} $ for long
times. As we have discussed, the $n^{5/3}$ behavior coincides with the
asymptotic MSD estimated numerically for a periodic polygonal channel
made of parallel walls which form angles of $90$ degrees
\cite{JeRo06,JBR08} for which one would naively expect ballistic
behaviour.

It seems there do not exist too many models, neither in terms of
deterministic nor stochastic dynamics, that exhibit sub-, super- and
normal diffusive regimes under single parameter variation. The slicer
map adds a new facet to this rather rare collection, as it generates
all these different types of diffusion in a strictly deterministic and
non-chaotic way. This suggests a mechanism explaining why in computer
simulations of polygonal billiards so many different types of
diffusion have been observed under parameter variation
\cite{ArGuRe00,ARV03,SchSt06,JeRo06,SaLa06,JBR08}. It may also help to
explain the severe difficulties to model diffusion in such systems by
a single, sufficiently simple anomalous stochastic process: We have
argued that, depending on the value of its control parameter, the
slicer diffusion matches mathematically to what is generated by very
different classes of stochastic processes, which is in line with
findings for polygonal billiard diffusion.

It would be highly desirable to construct a simple stochastic
process reproducing the full range of the slicer diffusion. It would
also be important to extract a slicer map from a given polygonal
billiard starting from first principles. This would enable to check
whether a slicing mechanism similar to the one proposed here in terms
of a power law distribution of slicers is realistic. Correlation
functions for the slicer map need to be calculated in order to fully
appreciate its dynamics. And as the slicer map is analytically
tractable, our model invites to play around with variations of the
slicer idea for better understanding the origin of non-chaotic
diffusion.

\vskip 20pt \noindent
{\bf Acknowledgements:}

The authors are grateful to Raffaella Burioni for illuminating
remarks.  LR acknowledges funding from the European Research Council,
7th Framework Programme (FP7), ERC Grant Agreement no. 202680. The EC
is notresponsible for any use that might be made of the data appearing herein.
CG acknowledges financial support from the MIUR through FIRB project
``Stochastic processes in interacting particle systems: duality,
metastability and their applications'', grant n. RBFR10N90W and 
the Institut Henry Poincar\'e for hospitality during the trimester ``Disordered systems, random spatial processes and some applications''.

\section*{Bibliography} 

\end{document}